\documentclass[letterpaper,10pt,onecolumn]{article}

\usepackage{amsmath, amssymb}

\usepackage{graphicx, import, wrapfig, setspace, caption, pdflscape}

\usepackage[numbers]{natbib}

\usepackage{hyperref}
\hypersetup{
colorlinks=true,
linkcolor=black,
citecolor=blue,
urlcolor=black,
linktoc=allo
}

\usepackage{fullpage}
\usepackage{phd_notations}

\graphicspath{{figures_draft/}}

\title{Equilibrium states of a liquid bridge between flexible sheets}
\author{Mouad Boudina and Gwynn J. Elfring\\
\small Department of Mechanical Engineering \& Institute of Applied Mathematics\\
\small University of British Columbia,
Vancouver, BC V6T 1Z4, Canada
}
\date{\today}

\usepackage{etoolbox}
\appto\appendix{\counterwithin{equation}{section}}

\begin{document}
\maketitle

\begin{abstract}
We study equilibrium states of a drop between flexible sheets clamped on both ends.
Revisiting first the case of parallel sheets, we find multiple equilibria which we classify in a parameter space. In solution branching diagrams we identify hysteresis cycles and folds indicating abrupt transitions, yet not necessarily leading to channel collapse. Between nonparallel sheets, a drop can stay in equilibrium away from the ends even when the liquid is totally wetting, which is impossible between nonparallel rigid plates. We also show that nonparallel sheets can delay or prevent collapse in comparison to straight channels, suggesting thus a mechanism that protects slender structures inside micro-devices from damage by surface tension.
\end{abstract}

\section{Introduction}
A drop deposited between hydrophilic slender structures, such as fibers, sheets or pillars, forms a liquid bridge that can apply a sufficient capillary compression able to collapse the structures \citep{mastrangelo1993}. This wet adhesion, also known as stiction \citep{tas1996}, is common since condensation is inevitable in humid environments \citep{delrio2007, wang2009} or rinsing during fabrication, e.g. by etching in wet medium \citep{guckel1989}. While an evaporating liquid turns a forest of nanopillars into an array of helical bundles \citep{pokroy2009}, it sadly ruins actuators, transducers and sensors made of micrometric slender beams and plates \citep{mastrangelo1997}. It is important to understand the states of a liquid bridge between slender structures to find the critical parameter values that bring about collapse and anticipate stiction.

A pair of elastic fibers contracts when trapping a drop of a wetting liquid. If the fibers are soft or close enough, they collapse and coalesce over a span \citep{duprat2015}. Ribbons carrying a liquid bridge simultaneously bend and twist, and eventually coalesce \citep{legrain2016}. A stretchable sheet trapping a liquid can also stick to a substrate, though in some cases only partially as a dimple forms at the centre \citep{butler2019}.

Notwithstanding the collapse and damage in micro-devices, liquid in narrow gaps is desired in adhesive applications because it joins adjacent soft substrates \citep{wexler2014} and spheres \citep{butt2010} stronger than rigid ones. For example, the adhesive force scales as the wet area for soft spheres, whereas it scales only as the radius for rigid ones \citep{butt2010}. In the similar problem of a cantilever squeezing a drop, adhesion improves when liquid squeezes and spreads out \citep{kwon2008}. From this perspective, elasticity of pads in tree frogs and set\ae\ in dock beetles is a beneficial asset \citep{gilet2019}. 

Between a pair of structures fixed at one end but with free tips, a liquid bridge forms in presence of pinning factors such as contact angle hysteresis \citep{bradley2021} or surface texture. An illustration of the latter is the fibers of a Chinese paint brush patterned with micro-metric squam\ae\ \citep{wang2014}. The fiber diameter and inter-fiber distance determine a particular range of heights where ink drops would stay pinned. Another example is the fibrous barbules of sandgrouse feathers. After soaking the belly feathers with water, liquid is retained in an array of barbules and stays trapped during the entire bird flight to their nests \citep{cade1967, mueller2023}. When these pinning factors are absent, the drop simply slides toward the tips while pulling the pair \citep{bradley2019}, and can even lead to coalescence as in goose feathers contaminated with oil \citep{duprat2012}. If the drop is initially trapped at the fixed end, the pair can take multiple deformation profiles for the same liquid volume, with the tips being open, in contact or coalescing \citep{taroni2012}.

The state of the liquid bridge between elastic structures depends on the relative bending stiffness, the volume of the drop, but also on the path taken through this parameter space. When both ends are fixed, a shrinking drop pulls a pair of sheets together steadily until contact at a particular volume \citep{chang2014}. Conversely, starting from an empty channel and increasing the liquid volume, a drop can abruptly collapse the channel with the sheets remaining in contact as the volume increases, and reopening at a volume higher than the one that led to contact for the shrinking drop. This hysteresis effect is commonly observed in adhesion problems \citep{mastrangelo1997, goryacheva2001, kwon2008, gilet2019, duprat2015, butler2019, siefert2022, chen2022}, meaning that a drop between doubly-clamped sheets is a system that can admit multiple equilibria for the same volume, which we shall determine explicitly.

In contrast to in a straight rigid channel, a drop cannot be in equilibrium between nonparallel plates unless pinned at the end. In fact, it moves toward the wedge \citep{reyssat2014}, even against gravity \citep{heng2015}, unless a pinning factor stabilizes it. A drop of a partially wetting liquid can remain in equilibrium thanks to contact angle hysteresis, but the angle between the walls has to be lower than a critical value \citep{ataei2017c}. Channel geometry is also a stabilizing factor, where smooth channels with a quadratically widening gap can suspend drops of a totally wetting liquid against gravity \citep{renvoise2009}. Wall flexibility is another factor that sustains drops in a tapered channel, which passively contracts such that the capillary pressure is equal on both menisci and cancels any cause of motion.

In the present work we study equilibria of a liquid bridge between flexible walls fixed at both ends. We first solve the special case of parallel sheets, describe the branching diagrams and classify the states into a parameter space. Then, we treat a liquid bridge between nonparallel sheets, and show that a drop can stay in equilibrium away from the ends even when the liquid is totally wetting, and that tilting the walls can prevent channel collapse.

\section{Problem formulation}
\begin{figure}
\centering
\includegraphics[width=\textwidth]{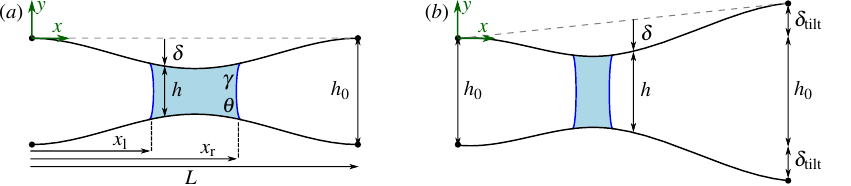}
\caption{Schematics of a drop between ($a$) parallel and ($b$) diverging flexible sheets. Note that the deflection~$\delta$ in ($b$) is relative to the original configuration in grey dashed line.}
\label{fig:scheme}
\end{figure}

We consider a drop of Newtonian liquid between two flexible sheets of length $L$ and bending modulus per unit width $B$, as shown in Figure \ref{fig:scheme}. The sheets are clamped on both sides and separated by a gap $h_{0}$ that is much smaller than $L$. We define the ratio $\epsilon = h_{0}/L \ll 1$. The contact angle between the liquid and the sheets is $\theta$, and we assume that the advancing and receding angles are equal. The system is symmetric with respect to the midline of the channel, hence we will focus on the deformation of the upper sheet alone. Let $\delta(x)$ denote the deflection of the sheets from the horizontal at abscissa $x$. The expression of the gap is
\begin{equation}
h(x) = h_{0} + 2[\deltatilt x/L + \delta(x)].
\label{eq:gap}
\end{equation} 
Note that the $y$-axis points upward, therefore a negative $\delta$ means a contraction of the channel. Throughout this problem we neglect gravity.

The liquid exerts on the sheets a capillary pressure $p$ in between $\xl$ and $\xr$, which are the positions of the left and right menisci of the drop respectively. Given that the channel aspect ratio is small, we assume the deflection satisfies the Euler-Bernoulli linear equation \citep{landau1970}. Taking the atmospheric pressure as a reference, we have
\begin{equation}
B\parDer{^{4}\delta}{x^{4}} = p,
\end{equation}
with $p = 0$ in the dry domain. Clamped sheets have a fixed position and direction on both sides, therefore
\begin{equation}
\delta = 0,\quad \delta' = 0,\quad \text{ at } x=0,L,
\label{eq:bc_clamped}
\end{equation}
where the prime $(.)'$ denotes the derivative with respect to $x$. At meniscus, $\xl$ and  $\xr$, the deflection $\delta$ and slope angle $\delta'$ are continuous. In absence of external torques, the second derivative $\delta''$ is also continuous. However, the third derivative $\delta'''$ is discontinuous owing the downward contact line force $-\gamma \sin(\theta \pm \delta')$. But since the gap is narrow $\epsilon = h_{0}/L \ll 1$, the force due to the contact line is negligible compared to the force owing to the capillary pressure $\gamma \ll \gamma (\xr-\xl)/h_{0} \sim \gamma L/h_{0}$, therefore we consider the third derivative also continuous.

We non-dimensionalize $x$ by $L$, $h$ and $\delta$ by $h_{0}$, and $p$ by the Young-Laplace pressure for an undeformed channel $2\gamma\cos\theta/h_{0}$. Writing dimensionless variables with a tilde, the gap height
\begin{equation}
\htld(\xtld) = 1 + 2[\deltatilttld\xtld + \deltatld(\xtld)],
\end{equation} 
is determined by
\begin{equation}
\parDer{^{4}\deltatld}{\xtld^{4}} =
\frac{2\ptld}{\calNec},
\label{eq:main}
\end{equation}
where $\calNec$ is the elastocapillary number \citep{mastrangelo1993}
\begin{align}
\calNec = \frac{Bh_{0}^{2}}{\gamma \cos\theta L^4},
\end{align}
which may be viewed as the ratio of length scales $\calNec = (\lec/L)^4$ where $\lec = (Bh_{0}^{2}/\gamma \cos\theta)^{1/4}$ is the elastocapillary length which balances the bending energy with the surface energy, $B\lec(h_{0}/\lec^{2})^{2} \sim \gamma\lec$.

The dimensionless boundary and continuity conditions are
\begin{subequations}
\begin{gather}
\deltatld|_{\xtld=0,1} = 0,\quad
\deltatld'|_{\xtld=0,1} = 0,\quad
\label{eq:bcs_01} \\
\deltatld|_{\xtld=\xtld\subrm{l,r}^{-}} = \deltatld|_{\xtld=\xtld\subrm{l,r}^{+}},\quad
\deltatld'|_{\xtld=\xtld\subrm{l,r}^{-}} = \deltatld'|_{\xtld=\xtld\subrm{l,r}^{+}},\quad
\deltatld''|_{\xtld=\xtld\subrm{l,r}^{-}} = \deltatld''|_{\xtld=\xtld\subrm{l,r}^{+}},\quad
\deltatld'''|_{\xtld=\xtld\subrm{l,r}^{-}} = \deltatld'''|_{\xtld=\xtld\subrm{l,r}^{+}}.
\label{eq:bcs_continuity}
\end{gather}
\label{eq:bcs}
\end{subequations}
where for brevity the notation $\xtld\subrm{l,r}$ refers to both $\xltld$ and $\xrtld$ simultaneously, while the $+$ and $-$ signs in the superscript indicate the left and right limit values respectively at these points. 

In equilibrium state the liquid is still, hence the Young-Laplace pressure in both left and right menisci must be equal
\begin{equation}
-\frac{\cos(\theta-\epsilon\deltatilttld-\epsilon\deltatld|'_{\xltld})}{\hxltld\cos\theta}
= -\frac{\cos(\theta+\epsilon\deltatilttld+\epsilon\deltatld|'_{\xrtld})}{\hxrtld\cos\theta}.
\end{equation}
To the leading order, the equilibrium condition reduces to the equality of gaps, $\hxltld = \hxrtld$,
\begin{equation}
\deltatilttld\xltld + \deltaxltld = \deltatilttld\xrtld + \deltaxrtld.
\label{eq:equilibrium_conditions}
\end{equation}

The pressure in the wet domain ($\xtld \in [\xltld,\xrtld]$) may be written 
\begin{align}
\ptld = -\frac{1}{\hxltld},\quad 
\end{align}
\label{eq:p_wet_dry}
while $p=0$ in the dry domains.

We integrate \eqref{eq:main} with pressures \eqref{eq:p_wet_dry} and find the analytical expression for the deflection $\deltatld$. To determine the twelve constants of integration and unknown positions $\xltld$ and $\xrtld$, we apply \eqref{eq:bcs} and \eqref{eq:equilibrium_conditions}, and close the system of equations by
fixing the (dimensionless) volume
\begin{equation}
\Vtld = \frac{V}{h_{0}L} = \int_{\xltld}^{\xrtld}\htld(\xtld)\rmd\xtld.
\end{equation}

For a given elastocapillary number $\calNec$, we solve the algebraic system above for the deflection. Instead of fixing the drop volume, we found it simpler to treat $\deltatld$ and $\Vtld$ as functions of $\xltld$ that we then vary. We relegate algebraic forms of the solutions to the Appendix and focus in the main text on physical description. In general, we find multiple equilibria for some ranges of drop volumes, which we illustrate by plotting the minimum gap $\hmintld = \min_{\xtld}\htld$ versus the volume, calculated by solving $\htld'=0$ subject to $\htld''>0$.

We distinguish three regimes: sheets that are open, in contact at a single point, and coalescing over a finite length. In the latter, the two detachment points are additional unknowns that we find by specifying a zero gap, slope and curvature. The nullity of the curvature, including in the detachment points, comes from the physical assumption that a very thin film intercalates the sheets in the coalesced region, which prevents solid-solid contact and adhesion work \citep{kim2006}, thus there is no mechanism to balance any internal moment. Otherwise the presence of an adhesive work due to solid-solid contact manifests as a bending moment concentrated at the detachment point and a discontinuity of the curvature \citep{plaut2001a}.

\section{Parallel sheets}
In the case of parallel sheets, $\deltatilttld=0$, we assume even symmetry about the midpoint $\xtld=1/2$,
\begin{subequations}
\begin{gather}
\xrtld = 1-\xltld,\\
\deltaxrtld = \deltaxltld,\quad
\deltaxrtld' = -\deltaxltld.
\end{gather}
\end{subequations} 
We will relax this assumption later in the general case of tilted sheets, and show that indeed only symmetric solutions exist in the case of parallel sheets with finite stiffness ($\calNec<\infty$).

\begin{figure}
\centering
\includegraphics[width=\textwidth]{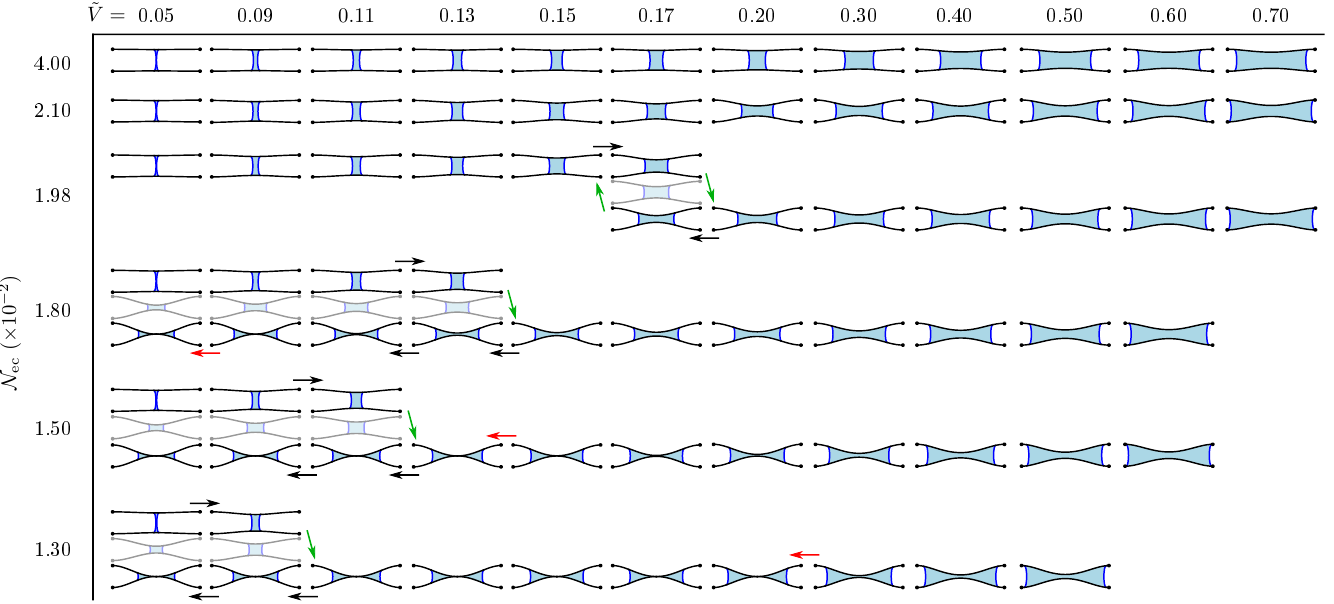}
\caption{Examples of deformation profiles in the case of parallel sheets, $\deltatilttld=0$, for different elastocapillary numbers $\calNec$. Transparent colours are unstable solutions. Red and green arrows indicate the steady and abrupt transitions. Black arrows mark the hysteresis loop.}
\label{fig:chart}
\end{figure}

\begin{figure}
\centering
\includegraphics[width=0.6\textwidth]{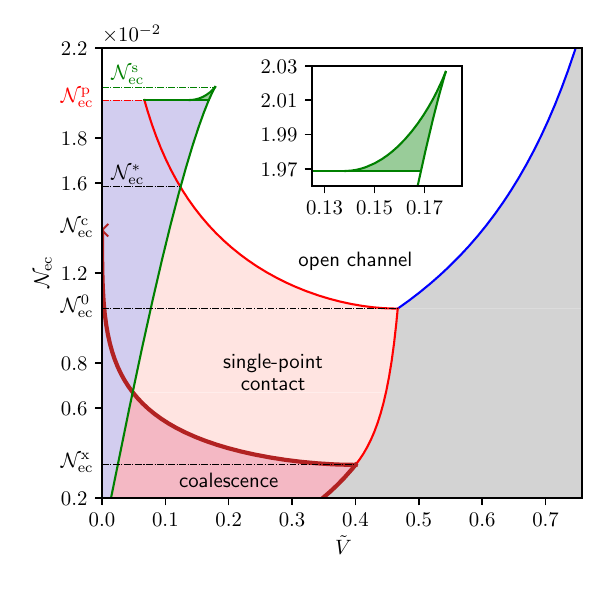}
\caption{Parameter space of possible solutions. The white regions indicate open channels, light red are channels closed with a single point of contact, while dark red indicates channels coalesced over a finite region; in all these areas there is only one stable solution. The grey region has no solutions while the purple and green regions have multiple stable solutions. The blue line indicates the maximum drop volume $\Vmaxtld$ that an open channel can carry. The green and red lines mark the snap and pinch transitions below thresholds $\calNecs$ and $\calNecp$ respectively. The cusped green area between $\calNecs$ and $\calNecp$ contains the hysteresis loop with three solutions in the open state. Below $\calNec^{*}$, the region in light red contains only the solution of sheets in contact at a single point. The blue and red lines intersect at $\calNec^{0}$ and $\Vmaxtld^{0}$, which represents a channel completely filled and collapsed at the same time. The maximum volume for a single-point contact state, between $\calNec^{0}$ and $\calNecx$ is given by $\Vmaxtld\suprm{c}$ drawn in red line. The dark red bold line between $\calNecc$ and $\calNecx$ marks the coalescence transition $\deltatld''|_{\xtld=1/2}=0$. Below $\calNecx$, sheets are always coalescing whereby the maximum volume is $\Vmaxtld$, also in bold red line.}
\label{fig:diagram_big}
\end{figure}

Examples of deformation profiles are shown in Figure~\ref{fig:chart}. The minimum volume is when the channel is empty, $\xltld=1/2$ and $\Vtld=0$, whereas the maximum volume occurs when the drop spans the whole channel, $\xltld=0$ and
\begin{equation}
\Vmaxtld = 1 - \frac{1}{180\calNec}.
\end{equation}
When the channel is full, only one solution exists across all elastocapillary numbers where the channel remains open. However, as the volume decreases, multiple stable and unstable equilibria are possible depending on the elasticity. The channel can pinch closed at a single point in the middle, or collapse over a finite length for lower elastocapillary numbers. In Figure \ref{fig:diagram_big} we show the parameter space for each of the different channel states with separatrix determined from our analytical solutions. The white regions indicate open channels, light red are channels closed with a single point of contact, while dark red indicates channels coalesced over a finite region; in all these areas there is only one stable solution. The grey region has no solutions while the purple and green regions have multiple stable solutions.

The parameter space in Figure~\ref{fig:diagram_big} reveals a global view of transitions between states that were only partially reported before, as we group geometric and structural parameters into a single dimensionless quantity, $\calNec$, and vary the drop volume as a parameter. Physically, we can freely vary the volume in infinitely wide sheets, unlike in fibers where drops must be small enough to avoid detachment over time \cite{aziz2019}.

\begin{figure}[t!]
\centering
\includegraphics[width=\textwidth]{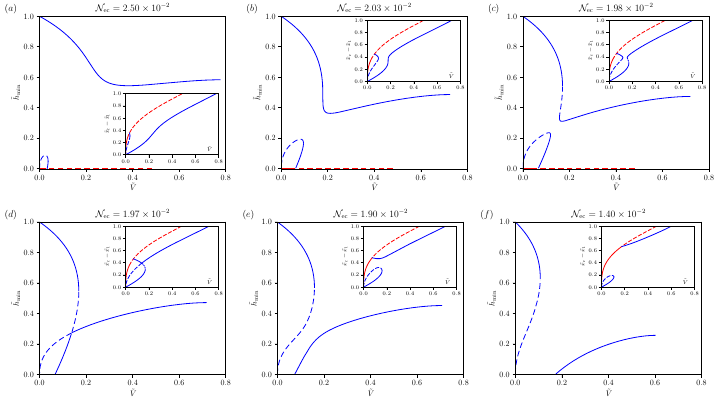}
\caption{Minimum gap and drop length (insets) versus volume for elastocapillary numbers. The red lines are the solutions of the collapsed state. Dashed lines indicate an unstable branch.}
\label{fig:hmin_vs_V}
\end{figure}

To explore the space of equilibrium shapes, we plot in Figure~\ref{fig:hmin_vs_V} the minimum gap $\hmintld$ and drop length $\xrtld-\xltld$ (insets) versus the drop volume $\Vtld$ for different elastocapillary numbers $\calNec$. There is only one equilibrium solution when the channel is full, $\Vtld=\Vmaxtld$, which we will refer to as the main branch. When the volume of the drop decreases, the menisci move away from the ends and the channel contracts in order to balance the high pressure load. Below a certain volume a second solution arises with the channel collapsed, however is unstable ($\hmintld=0$, red dashed line). Decreasing the volume further yields more equilibria.

Above the elastocapillary number $\calNecs \approx 2.03 \times 10^{-2}$, as the volume decreases below a certain value, the wet area becomes too small to sustain the channel deformation and $\hmintld$ relaxes back to 1 as the drop shrinks $\Vtld \rightarrow 0$ (Figure~\ref{fig:hmin_vs_V}$a$). At $\calNecs$, the main branch exhibits a hysteresis point \citep[p. 51]{seydel1988} (Figure~\ref{fig:hmin_vs_V}$b$). The number $\calNecs$ marks, therefore, the onset of mechanical instability, indicated by a cusp in the parameter space (Figure~\ref{fig:diagram_big}). Below, the main branch has two turning points (Figure~\ref{fig:hmin_vs_V}$c$) found by solving $\partial_{\xltld}\Vtld = 0$, and the problem admits three solutions in between. Since one-dimensional systems exchange stability at turning points \citep[p. 19]{iooss1990}, \citep[p. 63]{thompson1984}, \citep{thompson1979}, the middle branch is unstable and the transition between stable solutions is discontinuous, a type commonly referred to as a snap-through transition.

At $\calNecp \approx 1.97\times10^{-2}$, the main branch and a secondary branch meet at a transcritical point (Figure~\ref{fig:hmin_vs_V}$d$) \citep{seydel1988}. When $\calNec < \calNecp$, a filled channel pinches closed, $\hmintld=0$, without undergoing instability when the volume decreases down to a pinch volume (red line, Figure~\ref{fig:diagram_big}). In the words of \citet{chang2014}, this transition corresponds to an `equilibrium contact'. On the other hand, an empty channel, $\hmintld=1$, $\Vtld=0$, abruptly contracts as the volume increases up to a snap volume  (green line, Figure~\ref{fig:diagram_big}) but stays open, since the snap volume is larger than the pinch volume. Hence, a snap transition does not necessarily lead to the collapse of the pair as it might be understood \citep{chang2014, duprat2015}.

For elastocapillary numbers below $\calNec^{*} \approx 1.58\times10^{-2}$, nonetheless, the snap volume is smaller than the pinch volume (Figure~\ref{fig:hmin_vs_V}$f$). Starting from $\Vmaxtld$ and decreasing, the sheets pinch closed and stay so, as experimentally observed for a liquid between soft substrates \citep{wexler2014}. Conversely, from the point $\hmintld=1$, $\Vtld=0$ onwards, the channel abruptly collapses at the snap volume, remains so till the pinch volume then steadily reopens. The example in \citet{chang2014} and \citet{chen2022}, obtained analytically and numerically, corresponds then to $\calNec < \calNec^{*}$.

Further decreasing the elastocapillary number, the lower branch shifts down and shortens until vanishing at $\calNec^{0} = 1/96 \approx 1.04\times10^{-2}$, where the channel is entirely filled and collapsed at the same time. If $\calNec < \calNec^{0}$, the channel can be in an open state only for small volumes.

\begin{figure}
\centering
\includegraphics[width=\textwidth]{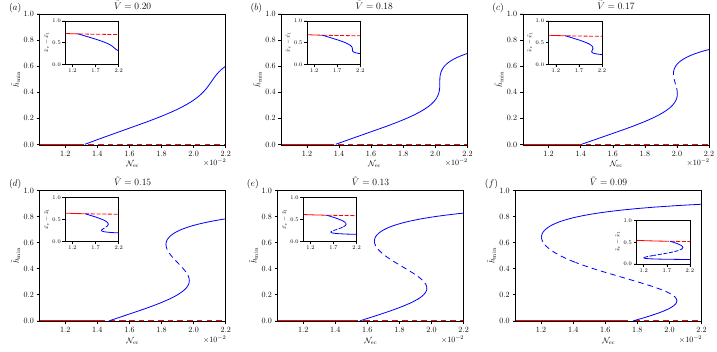}
\caption{Minimum gap and drop length (insets) versus elastocapillary length for different drop volumes. Same line and color code as in Figure~\ref{fig:hmin_vs_V}. After $\Vtld \approx 0.18$ we see the onset of hysteresis.}
\label{fig:hmin_vs_Nec}
\end{figure}

Another way to look at the equilibria is to fix the volume and vary the elastocapillary number as depicted in Figure~\ref{fig:hmin_vs_Nec}. The branching diagrams show three modes of collapse. The first mode is of a channel that steadily contracts as the elastocapillary number reduces (Figure~\ref{fig:hmin_vs_Nec}$a$). Below the hysteresis bifurcation at $\Vtld \approx 0.18$ (Figure~\ref{fig:hmin_vs_Nec}$b$), the second mode is of a channel that abruptly contracts but stays open, then steadily closes (Figure~\ref{fig:hmin_vs_Nec}$c,d,e$). Finally, the third mode is of a channel that abruptly collapses (Figure~\ref{fig:hmin_vs_Nec}$f$).

In the experiments of \citet{duprat2015}, a drop of constant volume was applied between two fixed parallel fibers. When the inter-distance decreased, equivalent to a reduction of $\calNec$ in our case, the fibers collapsed quickly at a certain value. They called this event a `zipping transition', after which the fibers were reported to be `almost in contact' after a `rapid change in the shape'. They also found a slope of the upper branch of the hysteresis cycle more inclined than the lower branch, similar to the one in the inset of Figure~\ref{fig:hmin_vs_Nec}($e$). On the other hand, \citet{siefert2022} carried out experiments where a pair of fibers were dipped into a liquid bath. By varying the fiber inter-distance at the origin and the immersed length, that is, varying $\calNec$ in our notation, they obtained a branching diagram that displays a turning point with stability exchange and a horizontal branch of the collapsed state, like the one of Figure~\ref{fig:hmin_vs_Nec}$(f)$.

When flexible enough, the sheets can contact one another and close the channel. The contact may occur at a single point at the center or over a finite length. Our analytical solutions allow us to disambiguate between these two cases by noting that a single point of contact requires positive curvature at the point of contact
\begin{equation}
\deltatld''|_{\xtld=1/2} > 0,
\end{equation}
which also implies a nonzero bending moment. This condition is always satisfied for $\calNec > \calNecc = 1/72 \approx 1.39\times10^{-2}$. For loose sheets of $\calNec \le \calNecc$, the liquid load can be sufficiently high that the sheets no longer stick at a single point but coalesce over a length and $\deltatld''|_{\xtld=1/2}=0$ (dark red, Figure~\ref{fig:diagram_big}). Finally, for very soft channels of $\calNec < \calNecx = 1/288 \approx 3.47\times10^{-3}$, inequality $\deltatld|''_{\xtld=1/2}>0$ is never satisfied, hednce the sheets always coalesce over a nonzero span.

\section{Diverging sheets}

\begin{figure}
\centering
\includegraphics[width=0.8\textwidth]{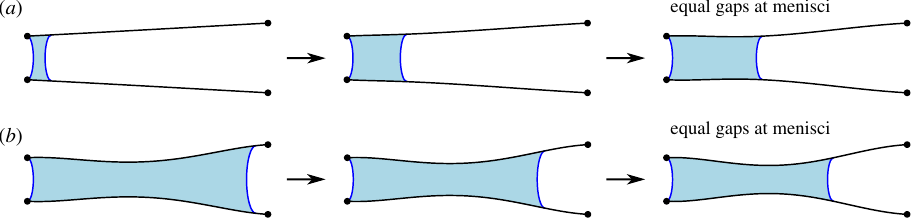}
\caption{Profiles of a drop pinned at the fixed end. ($a$) Increasing or ($b$) decreasing the volume moves the menisci away or toward each other until they become of equal size.}
\label{fig:pinned}
\end{figure} 

Now we consider a diverging channel, $\deltatilttld>0$, as sketched in Figure~\ref{fig:scheme}($b$). If the walls are rigid, the menisci would have different sizes and the drop would necessarily slide and pin at the edge, $\xltld = 0$ \citep{reyssat2014}. The contact angle~$\thetap$ at the pinned end is determined by setting the pressures equal on both menisci and taking the leading order,
\begin{equation}
\cos\thetap \approx \frac{\cos\theta}{\hxrtld}.
\end{equation}
For flexible sheets, the capillary pressure can contract the channel such that the meniscus on the left and right have equal size, $\hxltld = \hxrtld$, leading to an equilibrium state despite the asymmetry of the tapered configuration.

Starting from a small pinned drop at the edge and adding liquid (Figure~\ref{fig:pinned}$a$), the left meniscus shifts forward and expands, i.e. $\xrtld$ and $\hxrtld$ increase. After covering a certain area, the liquid load overtakes the bending force, the channel contracts more and the right meniscus gets smaller until equalling the size of the left meniscus, i.e. $\hxrtld = 1$. At this stage, the pinning angle equals the contact angle $\theta=\thetap$, and the solution is identical to the one found in the general case solving for the equality of gaps, $\hxltld=\hxrtld$. Similarly, starting from a drop filling the channel, $\xrtld=1$, and decreasing the volume (Figure~\ref{fig:pinned}$b$), the right meniscus shrinks until $\hxrtld=\hxltld=1$. By solving $\hxrtld=1$, we find the points connecting the branches of the pinned and sustained states, and deduce a limit elastocapillary number, $\calNeclim = (5.33\times10^{-3})/\deltatilttld$, above which solutions of equal menisci are absent, i.e. the drop always slides and pins to the wedge.

\begin{figure}
\centering
\includegraphics[width=\textwidth]{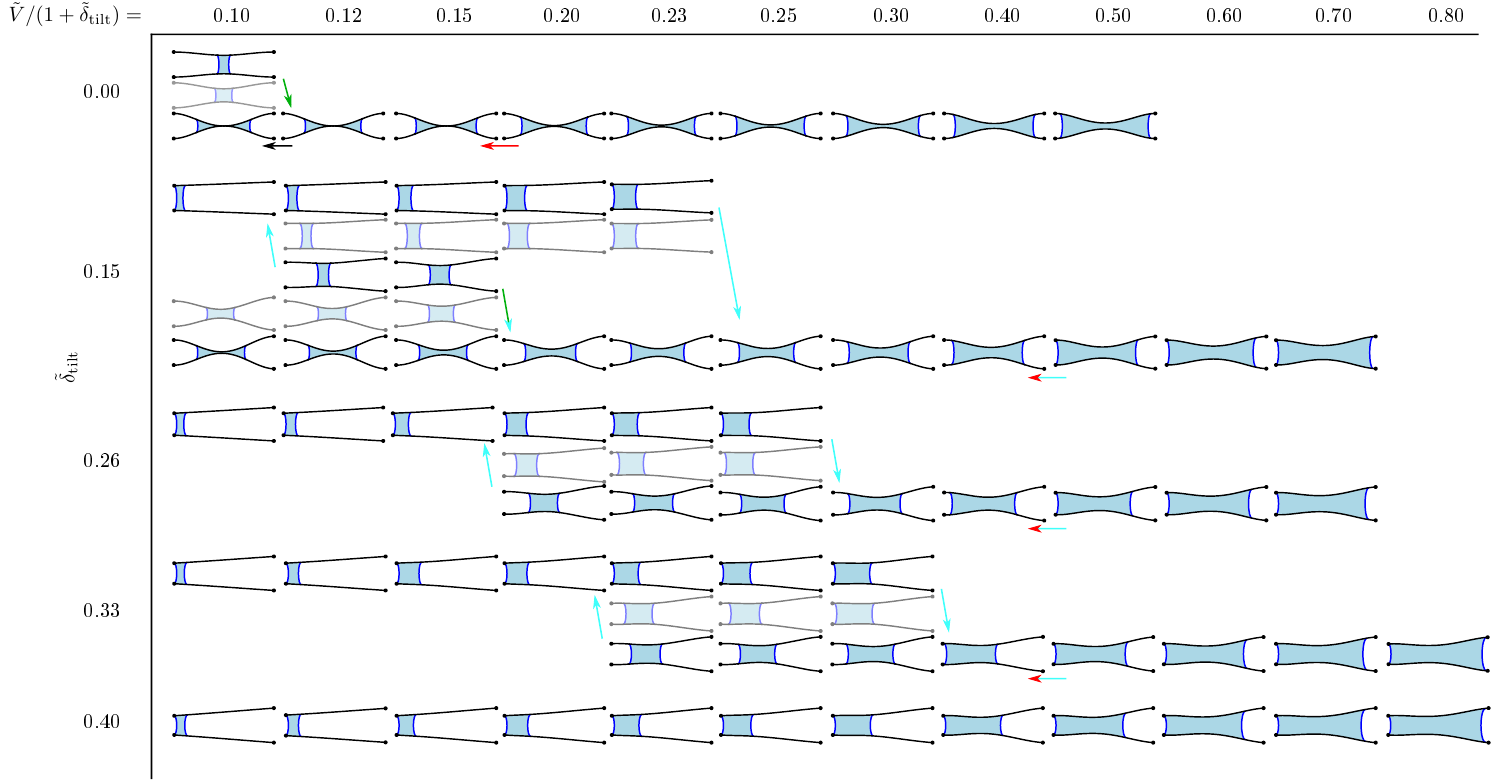}
\caption{Examples of deformation profiles in the case of diverging channels having $\calNec = 1.40\times10^{-2}$. Transparent colours are unstable solutions. Red and green arrows indicate the steady and abrupt transitions, cyan arrows mark a transition to or from a pinned solution.}
\label{fig:chart_tilt}
\end{figure}

\begin{figure}
\centering
\includegraphics[width=\textwidth]{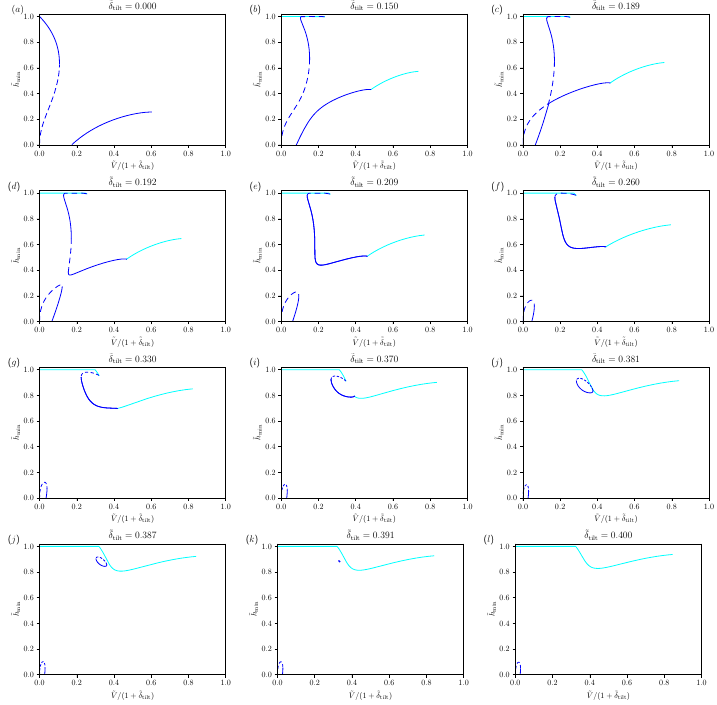}
\caption{Minimum gap versus normalized volume at different tilt angles for $\calNec=1.40\times10^{-2}$. Since the dimensionless volume can be greater than unity in a diverging channel, we normalize it by its maximum value for rigid walls, $\Vtld/(1+\deltatilttld)$. Light blue branches are the pinned state, and blue branches are the state with unpinned menisci.}
\label{fig:hmin_vs_V_Nec1.40}
\end{figure}

For $\calNec = 1.40\times10^{-2}$ we show examples of deformation profiles (Figure~\ref{fig:chart_tilt}) and branching diagrams (Figure~\ref{fig:hmin_vs_V_Nec1.40}). With this elasticity, the channel collapses whether initially empty or full when straight (Figure~\ref{fig:hmin_vs_V_Nec1.40}$a$), but only if initially full when slightly tapered (Figure~\ref{fig:hmin_vs_V_Nec1.40}$b,c$). Starting from the point $\Vtld=0$, $\hmintld=1$ and increasing the volume, the pinned drop stretches until the menisci have equal sizes, after which the system undergoes a snap transition down to the lower branch. Again, the channel suddenly contracts but remains open.

The branching diagram undergoes a transcritical bifurcation at $\deltatilttld \approx 0.189$ (Figure~\ref{fig:hmin_vs_V_Nec1.40}$c$), which is the minimum angle to prevent collapse, and a hysteresis bifurcation at $\deltatilttld \approx 0.209$ (Figure~\ref{fig:hmin_vs_V_Nec1.40}$e$). As the tilt increases, the amplitude of the jump between branches reduces, until $\deltatiltlimtld = (5.33\times10^{-3})/\calNec \approx 0.381$ where another transcritical point forms, connecting the branches of the pinned state and isolating the ones of the sustained state (Figure~\ref{fig:hmin_vs_V_Nec1.40}$i$). The latter becomes unreachable by a simple span of the volume from $\Vmaxtld$ to zero or opposite. It can be reached only by applying a large disturbance to the pinned state in the appropriate volume range \citep[p. 283]{bazant1991}, or if the system starts exactly from that equilibrium. Increasing further the tilt angle brings closer the turning points of the isolated branch that shrinks into an isola at $\deltatilttld \approx 0.391$ (Figure~\ref{fig:hmin_vs_V_Nec1.40}$k$) \citep[p. 84]{seydel1988}, so far unreported in elastocapillary problems. Isola centers have been observed in catalytic reactor models \cite{hlavacek1970, bizon2017}, liquid and granular flows \cite{kistler1994, alam2005} and fluid-structure interactions \cite{li2023}.

\section{Conclusions}
We present equilibrium solutions of a drop between sheets clamped on both ends and comprehensively elucidate the possible solutions in the parameter space of sheet flexibility and drop volume. This system exhibits hysteresis as observed in elastocapillary experiments \citep{fargette2014,siefert2022} and simulations \citep{chen2022}. In terms of practical experimental application, this parameter space can give measurement alternatives. For example, a particular transition in the equilibrium state for a given volume implies a corresponding elastocapillary number from which a physical quantity can be deduced, as done for the capillary pressure \citep{tas2010}. Similarly, if surface tension and Young’s modulus are known, a drop volume can be calibrated by decreasing the channel gap and marking a transition. This method avoids inferring the volume from the minimum gap through the relation $\hmintld$ versus $\Vtld$ as the channel deflection is usually difficult to measure with accuracy. On the other hand, the pair of slender sheets can serve as a detector in micro-scale systems, e.g. volume excess or humidity level due to a leak \citep{fargette2014}.

We also find that the channel avoids collapse above a critical tilt angle either upon volume increase or decrease. Therefore, tilting micro-beams in small devices is a possible solution to prevent stiction and protect from damage. In presence of gravity, a rigid channel can sustain a drop if the cross-section is curved \cite{renvoise2009,wang2013a}. Between undulating fibers, a train of liquid bridges falls down and merges depending on their size while the flow shows interesting dynamics \cite{gabbard2023}. A worthwhile extension would study the role of gravity on walls that are intrinsically curved and flexible at the same time.

The present study treats a channel of a simple nonuniform cross-section. Other types of nonuniformities offer interesting alternatives as well, such as beams of intrinsic curvatures or heterogeneous Young's modulus like in the hair of bee tongues \cite{wei2022}. In a half-cut conical rod, for instance, a growing drop reorients to the curved side and merges with a neighbouring drop on another rod and so forth, until liquid collects and get directionally transported \cite{feng2020}. On the other hand, since a film of composite material can be more difficult to peel from a substrate \cite{yin2023}, there are reasons to believe that beams of heterogeneous elasticity might display different equilibria when trapping a liquid bridge. Additionally, the enhanced adhesive properties of mushroom-shaped contact elements and their practical advantages in medical technology and robotics \cite{heepe2014} incites analyses of a liquid bridge between substrates of different shapes and thicknesses.

Another valuable avenue would consider beams with a surface tension gradient or a biphasic drop of two immiscible liquids \cite{bico2002}. The latter takes inspiration from the substance that bees secrete to collect pollen grains, which is composed of a liquid with an oily phase attenuating adhesion variations with humidity \cite{shin2019}. The results might shed light on oil separation applications, already studied for rigid walls \cite{luo2014}.

\appendix
\renewcommand{\theequation}{A\arabic{equation}}
\setcounter{equation}{0}
\section{General case}
We first integrate \eqref{eq:main} in the dry domains $[0,\xltld]$ and $[\xrtld,1]$ and apply the boundary conditions \eqref{eq:bcs_01}. We find
\begin{subequations}
\begin{align}
&\deltatld(\xtld) = \deltaxltld \left( \frac{\xtld}{\xltld} \right)^{2} \left( 3 - 2\frac{\xtld}{\xltld} \right)
+ \xltld \deltaxltld'  \left(\frac{\xtld}{\xltld}\right)^{2} \left(\frac{\xtld}{\xltld} - 1 \right),
\quad \forall \xtld \in [0,\xltld],
\label{eq:deltadry1}\\
&\deltatld(\xtld) = \deltaxrtld \fracpow{1-\xtld}{1-\xrtld}{2} \left[ 3 - 2 \frac{1-\xtld}{1-\xrtld} \right] 
- (1-\xrtld)\deltaxrtld' \fracpow{1-\xtld}{1-\xrtld}{2} \left[ \frac{1-\xtld}{1-\xrtld} - 1 \right],
\quad \forall \xtld \in [\xrtld,1],
\label{eq:deltadry2}
\end{align}
\end{subequations}
from which we extract the relations
\begin{subequations}
\begin{gather}
\deltaxltld'' = -\frac{6\deltaxltld}{\xltld^{2}} + \frac{4\deltaxltld'}{\xltld}, \quad
\deltaxltld''' = -\frac{12\deltaxltld}{\xltld^{3}} + \frac{6\deltaxltld'}{\xltld^{2}},
\label{eq:deltaxl_pp_ppp}\\
\deltaxrtld'' = -\frac{6\deltaxrtld}{(1-\xrtld)^{2}} - \frac{4\deltaxrtld'}{1-\xrtld}, \quad
\deltaxrtld''' = \frac{12\deltaxrtld}{(1-\xrtld)^{3}} + \frac{6\deltaxrtld'}{(1-\xrtld)^{2}}.
\label{eq:deltaxr_pp_ppp}
\end{gather}
\end{subequations}

Then we integrate \eqref{eq:main} in the wet domain $[\xltld,\xrtld]$, and use \eqref{eq:deltaxl_pp_ppp} to get
\begin{equation}
\deltatld(\xtld) = \deltaxltld \left( \frac{\xtld}{\xltld} \right)^{2} \left( 3 - 2\frac{\xtld}{\xltld} \right)
+ \xltld \deltaxltld'  \left(\frac{\xtld}{\xltld}\right)^{2} \left(\frac{\xtld}{\xltld} - 1 \right)
- \frac{(\xtld-\xltld)^{4}}{12\calNec\hxltld}.
\label{eq:deltawet}
\end{equation}

Next, we evaluate the derivatives of \eqref{eq:deltawet} at $\xrtld$, and use the relations \eqref{eq:deltaxr_pp_ppp} in view of the continuity conditions \eqref{eq:bcs_continuity}. We obtain
\begin{subequations}
\begin{gather}
\deltaxltld = -\frac{\xltld^{2}(\xrtld-\xltld)P(\xltld,\xrtld)}{12\calNec\hxltld},\label{eq:deltaxl_general} \\
\deltaxltld' = -\frac{\xltld(\xrtld-\xltld)Q(\xltld,\xrtld)}{6\calNec\hxltld},\\
\deltaxrtld = -\frac{(\xrtld-\xltld)R(\xltld,\xrtld)}{12\calNec\hxltld},
\end{gather}
\end{subequations}
and $\deltaxrtld'$ is deduced from \eqref{eq:deltaxl_general} and \eqref{eq:deltawet}. Here $P$, $Q$ and $R$ are the following bivariate polynomials
\begin{subequations}
\begin{align}
P(\xrtld,\xltld) &=  -2\xltld^{4} + \xltld^{3}(-2\xrtld+7) + \xltld^{2}(-2\xrtld^{2}+7\xrtld- 8) + \xltld(-2\xrtld^{3}+7\xrtld^{2}-8\xrtld+2) + (3\xrtld^{3}-8\xrtld^{2}+6\xrtld),\label{eq:P}\\
Q(\xrtld,\xltld) &= -3\xltld^{4} + \xltld^{3} (- 3\xrtld + 9) + \xltld^{2}(- 3\xrtld^{2} + 9\xrtld - 8) + \xltld(-3\xrtld^{3} + 9\xrtld^{2} - 8\xrtld) + (3\xrtld^{3} - 8\xrtld^{2} + 6\xrtld),\\
R(\xltld,\xrtld) &= \xltld^{3}(-1-2\xrtld^{3}+3\xrtld^{2}) + \xltld^{2}(-2\xrtld^{4}+7\xrtld^{3}-8\xrtld^{2}+3\xrtld) + \xltld(-2\xrtld^{5}+7\xrtld^{4}-8\xrtld^{3}+3\xrtld^{2}) \nonumber\\ &\qquad+ (-2\xrtld^{6}+7\xrtld^{5}-8\xrtld^{4}+3\xrtld^{3}).\label{eq:R}
\end{align}
\end{subequations}

In particular, inserting $\hxltld = 1 + 2\deltatilttld\xltld+2\deltaxltld$ into \eqref{eq:deltaxl_general} gives
\begin{equation}
\hxltld = \frac{1}{2} \left[ 1+2\deltatilttld\xltld \pm \sqrt{(1+2\deltatilttld\xltld)^{2} - \frac{2\xltld^{2}(\xrtld-\xltld)P}{3\calNec}} \right],
\label{eq:hxl_general}
\end{equation}
and the expression of the volume $\Vtld = V/h_{0}L = \int_{\xltld}^{\xrtld}\htld(\xtld)\rmd \xtld$ is
\begin{equation}
\Vtld = \xrtld-\xltld + \deltatilttld(\xrtld^{2}-\xltld^{2}) - \frac{B(\xltld,\xrtld)}{30\calNec\hxltld},
\label{eq:volume_tilt}
\end{equation}
with
\begin{equation}
B(\xltld,\xrtld) = (\xrtld-\xltld)[(\xrtld-\xltld)^{4} + 10(\xrtld^{3}-\xltld^{3})(P/2-Q/3) + 5(\xrtld^{4}-\xltld^{4})(Q/2-P/2)].
\label{eq:B}
\end{equation}
Here we neglected the empty volume due to the meniscus curvature \cite{taroni2012}. Finally, we calculate the minimum gap $\hmintld$ by numerically solving the third-order equation $\htld'=0$, i.e. $\deltatilttld + \deltatld' = 0$,
\begin{equation}
\left(\deltatilttld + \frac{\xltld^{3}}{3\calNec\hxltld} \right)
+ \left( \frac{6\deltaxltld}{\xltld^{2}} - \frac{2\deltaxltld'}{\xltld} - \frac{\xltld^{2}}{\calNec\hxltld} \right) \xtld
+ \left( \frac{3\deltaxltld'}{\xltld^{2}} - \frac{6\deltaxltld}{\xltld^{3}} + \frac{\xltld}{\calNec\hxltld} \right) \xtld^{2}
- \frac{\xtld^{3}}{3\calNec\hxltld} = 0,
\label{eq:argmin}
\end{equation}
provided $\deltatld'' > 0$.

The equality of gaps \eqref{eq:equilibrium_conditions} reads as $\deltaxrtld-\deltaxltld = (-\deltatilttld)(\xrtld-\xltld)$, i.e.
\begin{equation}
\deltatilttld + \frac{\xltld^{2}P-R}{12\calNec\hxltld} = 0,
\end{equation}
and using \eqref{eq:hxl_general} becomes
\begin{equation}
\frac{\xltld^{2}P-R}{6\calNec(-\deltatilttld)} - (1+2\deltatilttld\xltld)= \pm \sqrt{(1+2\deltatilttld\xltld)^{2} - \frac{2\xltld^{2}(\xrtld-\xltld)P}{3\calNec}},
\label{eq:equal_gaps}
\end{equation}
which reduces to solving
\begin{equation}
\frac{(\xltld^{2}P-R)^{2}}{12\calNec\deltatilttld} + (1+2\deltatilttld\xltld)(\xltld^{2}P-R) + 2\deltatilttld\xltld^{2}(\xrtld-\xltld)P = 0.
\label{eq:equal_gaps_full}
\end{equation}
From \eqref{eq:P} and \eqref{eq:R} we get
\begin{equation}
\xltld^{2}P - R = (\xltld+\xrtld-1)(\xrtld-\xltld)
[\xrtld(1-\xrtld) + \xltld(1-\xltld)]
[2\xrtld(1-\xrtld) + 2\xltld(1-\xltld) + 1 - (1-\xltld)(1-\xrtld)].
\label{eq:2xP_R_factorized}
\end{equation}
In the general case, equation \eqref{eq:equal_gaps_full} is a twelveth-order polynomial in $\xltld$ which we solve for all $\xrtld$ such that $0 \le \xltld < \xrtld \le 1$. In the particular case of parallel sheets, $\deltatilttld=0$, we have $\xltld^{2}P-R=0$, hence \eqref{eq:2xP_R_factorized} implies that $\xltld+\xrtld=1$.

\section{Parallel case}
\renewcommand{\theequation}{B\arabic{equation}}
\setcounter{equation}{0}
\subsection{Open sheets}
Owing to symmetry, the gap at the meniscus becomes
\begin{equation}
\hxltld = \frac{1}{2} \left[ 1 \pm \sqrt{
1 - \frac{2\xltld^{2} (1-2\xltld) (1-2\xltld^{2})}{3\calNec}} \right],
\label{eq:hxl_open}
\end{equation}
the volume simplifies to
\begin{equation}
\Vtld = 1-2\xltld - \frac{(1-2\xltld)^{2}(1+4\xltld+12\xltld^{2}-8\xltld^{3}-20\xltld^{4})}{180\calNec\hxltld},\\
\label{eq:volume_open}
\end{equation}
and the minimum gap is the gap at the midpoint,
\begin{equation}
\hmintld =  \htld|_{\xtld=1/2} = 1 - \frac{\xltld^{4} - \xltld^{3} + 1/16}{6\calNec\hxltld}.
\label{eq:hmin_open}
\end{equation}

\subsubsection*{Snap transition}
From \eqref{eq:hxl_open} and \eqref{eq:volume_open}, the equation defining the snap transition, $\partial_{\xltld}\Vtld=0$, has the following expression
\begin{equation}
\begin{gathered}
2 - \frac{\xltld^{2}(1-2\xltld)(1-2\xltld^{2})}{\calNec}
+ \frac{\xltld(1-2\xltld)^{2}(1+\xltld-4\xltld^{2}+190\xltld^{3}-4\xltld^{4}-748\xltld^{5}+760\xltld^{7})}{270\calNec^{2}} \\
\pm 2\left[ 1 - \frac{2\xltld^{2}(1-2\xltld)(1-2\xltld^{2})}{3\calNec} \right] \sqrt{1 - \frac{2\xltld^{2}(1-2\xltld)(1-2\xltld^{2})}{3\calNec}} =0.
\end{gathered}
\label{eq:snap}
\end{equation}

\subsubsection*{Pinch transition}
Given \eqref{eq:hmin_open}, the equation defining the pinch transition, $\hmintld=0$, is
\begin{equation}
\frac{\xltld^{4}-\xltld^{3}+1/16}{3\calNec} - 1 = \pm \sqrt{1 - \frac{2\xltld^{2}(1-2\xltld)(1-2\xltld^{2})}{3\calNec}},
\label{eq:pinch}
\end{equation}
which we raise to the square to obtain an eighth-order polynomial equation
\begin{equation}
\xltld^{8} - 2\xltld^{7} + \xltld^{6} + 24\calNec \xltld^{5} + (1/8 - 18\calNec)\xltld^{4} - (1/8+6\calNec)\xltld^{3} + 6\calNec\xltld^{2} + (1/256 - 3\calNec/8) = 0.
\label{eq:pinch2}
\end{equation} 

\subsubsection*{Expression of $\calNec$ in terms of $\xltld$}
To draw the curves $\hmintld$ versus $\calNec$ in Figure~\ref{fig:hmin_vs_Nec}, we invert the expression of the volume in \eqref{eq:volume_open} and find
\begin{equation}
\calNec = \frac{(1-2\xltld)^{3}(1+4\xltld+12\xltld^{2}-8\xltld^{3}-20\xltld^{4})^{2}}
{180(1-2\xltld-\Vtld)
[(1-2\xltld)(1+4\xltld+12\xltld^{2}-8\xltld^{3}-20\xltld^{4}) - 30\xltld^{2}(1-2\xltld^{2})(1-2\xltld-\Vtld)]}.
\end{equation}

\subsection{Single-point contact}
To find the deformation profile of sheets with a single-point contact, we use the deflection expression \eqref{eq:deltawet} under the following boundary conditions
\begin{equation}
\deltatld|_{\xtld=1/2} = -\frac{1}{2},\quad
\deltatld'|_{\xtld=1/2} = 0,
\end{equation}
and obtain
\begin{equation}
\hxltld = \frac{(1-2\xltld)^{2}(1+4\xltld)}{2} \left[ 1 \pm \sqrt{1 - \frac{\xltld^{2}}{6\calNec(1+4\xltld)}} \right],
\label{eq:hxl_stick}
\end{equation}
and
\begin{equation}
\Vtld = \frac{(1-2\xltld)^{3}(1+2\xltld)}{2} - \frac{(1-2\xltld)^{5}(1+10\xltld+60\xltld^{2}+120\xltld^{3})}{2880\calNec\hxltld}.
\label{eq:volume_stick}
\end{equation}

The red horizontal lines $\hmintld=0$ versus $\calNec$ in Figure~\ref{fig:hmin_vs_Nec} are obtained by inverting \eqref{eq:volume_stick},
\begin{equation}
\calNec = \frac{(1+10\xltld+60\xltld^{2} + 120\xltld^{3})^{2}}
{1440(1+4\xltld)[1+2\xltld - 2\Vtld/(1-2\xltld)^{3}]
[1+10\xltld+60\xltld^{2}+120\xltld^{3} - 60\xltld^{2}[1+2\xltld - 2\Vtld/(1-2\xltld)^{3}]]
}.
\end{equation}

\subsection{Coalescing sheets}
The state of the sheets transitions from a single-point contact to a contact over a nonzero length when $\deltatld|''_{\xtld=1/2} = 0$. The explicit form of this condition is the following fourth-order polynomial equation
\begin{equation}
144\xltld^{4} + 96\xltld^{3} + 8(5-576\calNec)\xltld^{2} + 8(1-288\calNec)\xltld + (1-288\calNec) = 0.
\label{eq:coalesce_transit}
\end{equation}

The coalescence condition $\deltatld|''_{\xtld=\xctld}=0$ writes
\begin{equation}
\frac{(\xctld-\xltld)^{2}(\xctld^{2}+2\xltld\xctld+3\xltld^{2})}{9\calNec} = \frac{(\xctld-\xltld)^{2}(\xctld+2\xltld)}{\xctld^{3}} \left[ 1 \pm \sqrt{1 -  \frac{2\xltld^{2}\xctld^{3}}{3\calNec(\xctld+2\xltld)}} \right].
\label{eq:eq_coalesce}
\end{equation}
Keeping aside the case $\xctld=\xltld$ which gives a zero volume, equation \eqref{eq:eq_coalesce} simplifies into
\begin{equation}
\frac{\xltld^{4}\zetac^{3}(\zetac^{2}+2\zetac+3)}{9\calNec(\zetac+2)} - 1 = \pm \sqrt{1 - \frac{2\xltld^{4}\zetac^{3}}{3\calNec(\zetac+2)} },
\end{equation}
which admits solutions for all $\zetac = \xctld/\xltld \ge 1$ and $\calNec/\xltld^{4} \ge 2/9 \approx 0.22$. This equation reduces to solving the roots of the polynomial
\begin{equation}
\zetac^{6} + 4\zetac^{5} + 10\zetac^{4} + 12\zetac^{3}+ \left(9-\frac{18\calNec}{\xltld^{4}}\right)\zetac^{2} - \frac{72\calNec}{\xltld^{4}}\zetac - \frac{72\calNec}{\xltld^{4}} = 0.
\label{eq:eq_coalesce_poly}
\end{equation}
For every $\xltld \in [0, (9\calNec/2)^{1/4}]$, we solve \eqref{eq:eq_coalesce_poly} and keep the solutions such that $\xctld = \zetac\xltld \le 1/2$.

\section{Pinned drop}\label{app:pin}
\renewcommand{\theequation}{C\arabic{equation}}
\setcounter{equation}{0}
The deflection based on the coordinate $\xrtld$ is
\begin{equation}
\deltatld(\xtld) = \deltaxrtld \fracpow{1-\xtld}{1-\xrtld}{2} \left[ 3 - 2 \frac{1-\xtld}{1-\xrtld} \right] - (1-\xrtld)\deltaxrtld' \fracpow{1-\xtld}{1-\xrtld}{2} \left[ \frac{1-\xtld}{1-\xrtld} - 1 \right] - \frac{(\xrtld-\xtld)^{4}}{12\calNec\hxrtld}\calH(\xrtld-\xtld),
\label{eq:delta_based_xr}
\end{equation}
where $\calH$ is the Heaviside step function. The explicit form of equation $\hxrtld = 1$ is
\begin{equation}
1 - 2\deltatilttld\xrtld = \pm \sqrt{(1+2\deltatilttld\xltld)^{2} - \frac{2\xrtld^{4}(1-\xrtld)^{2}(3-2\xrtld)}{3\calNec}},
\label{eq:pin}
\end{equation}
which, excluding the trivial case $\xrtld=0$, reduces to
\begin{equation}
\calNec\deltatilttld = \frac{\xrtld^{3}(3 - 8\xrtld+ 7\xrtld^{2} - 2\xrtld^{3})}{12}.
\label{eq:pin_polynomial}
\end{equation}
The right-hand side of \eqref{eq:pin_polynomial} has a local maximum at $\xrtld=(23-\sqrt{97})/24$ giving $\calNec\deltatilttld \approx 5.3346\times10^{-3}$.

\bibliographystyle{unsrtnat}
\bibliography{phd_biblio}

\end{document}